\documentclass{PoS}

\usepackage{lineno}
\usepackage[locale=US, exponent-product=\cdot]{siunitx}
\usepackage{amsmath}
\usepackage{mathtools}
\usepackage{subcaption}
\usepackage{multirow}
\usepackage{setspace}

\title{Searches for steady neutrino emission from 3FHL blazars using eight years of IceCube data from the Northern hemisphere}

\ShortTitle{3FHL blazar stacking}

\author{
The IceCube Collaboration\footnote{For collaboration list, see PoS(ICRC2019) 1177.}\\
{\itshape \href{http://icecube.wisc.edu/collaboration/authors/icrc19_icecube}{http://icecube.wisc.edu/collaboration/authors/icrc19\_icecube}}\\
E-mail: \email{mhuber@icecube.wisc.edu}
}

\abstract{Located at the South Pole, the IceCube Neutrino Observatory is the world largest neutrino
telescope, instrumenting one cubic kilometre of Antarctic ice at a depth between \SI{1450}{\meter} to
\SI{2450}{\meter}. In 2013 IceCube reported the first observations of a diffuse astrophysical high-energy
neutrino flux. Although the IceCube Collaboration has identified more than 100 high-energy
neutrino events, the origin of this neutrino flux is still not known. Blazars, a subclass of Active
Galactic Nuclei and one of the most powerful classes of objects in the Universe, have long been
considered promising sources of high energy neutrinos. A blazar origin of this high-energy
neutrino flux can be examined using stacking methods testing the correlation between IceCube
neutrinos and catalogs of hypothesized sources. Here we present the results of a stacking analysis for 1301 blazars from the third catalog of hard \textit{Fermi}-LAT sources (3FHL). The analysis is performed on 8 years of
through-going muon data from the Northern Hemisphere, recorded by IceCube between 2009
and 2016. No excess of neutrinos from the blazar position was found and first limits on
the neutrino production of these sources will be shown \\

\vspace{4mm}
{\bfseries Corresponding authors:}
\speaker{Matthias Huber}$^{1}$\\
{$^1$ Technische Universit\"at M\"unchen, Physik-Departement, James-Franck-Str. 1, 85748 Garching}\\

}

\FullConference{36th International Cosmic Ray Conference -ICRC2019-\\
		July 24th - August 1st, 2019\\
		Madison, WI, U.S.A.}

\begin{document}

\section{Introduction}
\label{sec:intro}

With the first discovery of an astrophysical neutrino flux at PeV energies the IceCube collaboration marked the beginning of a new era of study not only in high energy neutrino
astronomy but more generally in the study of the high-energy Universe . Due to their neutral charge and their very low interaction cross sections neutrinos are neither deflected by magnetic fields nor significantly absorbed in the interstellar medium. Consequently, neutrinos carry unique information about their site of production. They are not only ideal messengers for the directional origin of the most powerful objects in the Universe but can also give insight in the acceleration mechanisms 
within these objects at the highest energies.

In September 2017 IceCube detected a high-energy neutrino event in spatial and temporal coincidence with a gamma-ray flare from the blazar TXS 0506+056 \cite{IceCube:2018dnn}. Subsequent studies revealed first evidence for neutrino emission from the direction of the blazar TXS 0506+056 in 2014/15 \cite{IceCube:2018cha}.

Blazars, a subclass of Active Galactic Nuclei (AGNs) hosting a jet of highly relativistic particles pointing towards Earth, are among the most luminous objects in the entire Universe \cite{Urry:1995mg}. Depending on the conditions for particle acceleration at the site of these objects, they can be regarded as potential extragalactic sources for the emission of high energy neutrinos \cite{Padovani:2014bha}.

In order to further investigate the possibility of neutrino production not only from the blazar TXS 0506+056, but more general from a population of blazars, we search for a steady neutrino signal from sources listed in the \textit{third Catalog of Hard Fermi-LAT Sources} (3FHL) \cite{TheFermi-LAT:2017pvy}. 
Using a source stacking method, this analysis studies 8 years of data from the IceCube Neutrino Observatory \cite{Aartsen:2016nxy}.
The IceCube Observatory is the world's largest neutrino telescope, instrumenting one cubic kilometre of Antarctic ice at a depth between \SI{1450}{\meter} to \SI{2450}{\meter}. 
The detection principle relies on the observation of Cherenkov light emitted from secondary particles created in interactions of neutrinos within the Antarctic ice or the nearby bedrock.
Muons, which produce a long luminous track within the detector, are most suitable to get an accurate pointing towards the origin of the events. 
Above energies of $\SI{10}{\tera\electronvolt}$ the direction of muon events in IceCube can be reconstructed with a median angular resolution of less than $\SI{1}{\degree}$ 
\cite{Aartsen:2016oji}. Hence the outcomes of this analysis are based on a sample of approximately $\SI{500000}{}$ through-going muon events from the Northern hemisphere integrated over a livetime of eight years.

\section{Blazar populations}
\label{sec:blazar}

Among the most luminous objects in the universe, blazars generate photons over a broad
emission band ranging from radio to \SI{}{\tera\electronvolt} energies. 
In general, the spectral energy distribution (SED) of blazars is composed of two broad humps.
While the low-energy peak between infrared and X-ray energies can be associated with 
synchrotron radiation from relativistic electrons, different potential scenarios exist for
the generation of the second hump at gamma-ray energies.

Based on different properties of the spectral energy distribution, blazars can be further
classified into flat-spectrum radio quasars (FSRQs) and BL Lacertae objects (BL Lacs). 
While the latter only show weak emission lines in the optical spectrum, strong and broad emission
lines are visible in the SED of FSRQs \cite{Urry:1995mg}. 
A second complementary classification is based on the non-thermal emission of blazars.
It makes use of the rest frame value of the frequency of the synchrotron peak $\nu_{peak}^S$,
reflecting the maximum energy of the accelerated electrons within the relativistic jets.
According to the position of the synchrotron peak, blazars are categorized as low and
intermediate synchrotron peaked (LSP and ISP) if $\nu_{peak}^S<\SI{e15}{\hertz} (\sim \SI{4}{\electronvolt})$ and high synchrotron peaked (HSP) otherwise. 

Theoretical models predict that neutrinos could be produced via the decay of charged
pions from photo-hadronic interactions of high energy protons with ambient photons or gas within 
the jets. This neutrino flux would be accompanied by a flux of very high energy (VHE) gamma-rays
from the decay of neutral pions. 

The existence of the highest-energy IceCube neutrinos ($\sim \SI{}{\peta\electronvolt}$) implies 
the generation of a gamma-ray flux in the $\geq \SI{}{\tera\electronvolt}$ range
\cite{PhysRevD.78.034013}. Hence, in general, blazars with higher gamma-ray flux are assumed to 
emit more high-energy neutrinos. Yet this
conclusion has to be treated carefully. Since VHE gamma-rays are strongly attenuated by extragalactic
background light (EBL) or reabsorbed at the source, the photon-neutrino connection on a source by source basis might be diluted.
Given that for the theoretical description of the low energy synchrotron hump the
existence of relativistic electrons is inevitable, it seems reasonable to assume that 
also the acceleration of hadrons is possible within the jets of blazars. Nevertheless, it has not
yet been possible to determine the fraction of the hadronic component to the high-energy hump 
of blazars nor to disprove purely leptonic models.
The detection of neutrinos from blazar populations would directly prove the existence of 
this hadronic component.

\begin{figure}[t]
\centering
\begin{subfigure}{.45\textwidth}
  \centering
  \includegraphics[width=1.\linewidth]{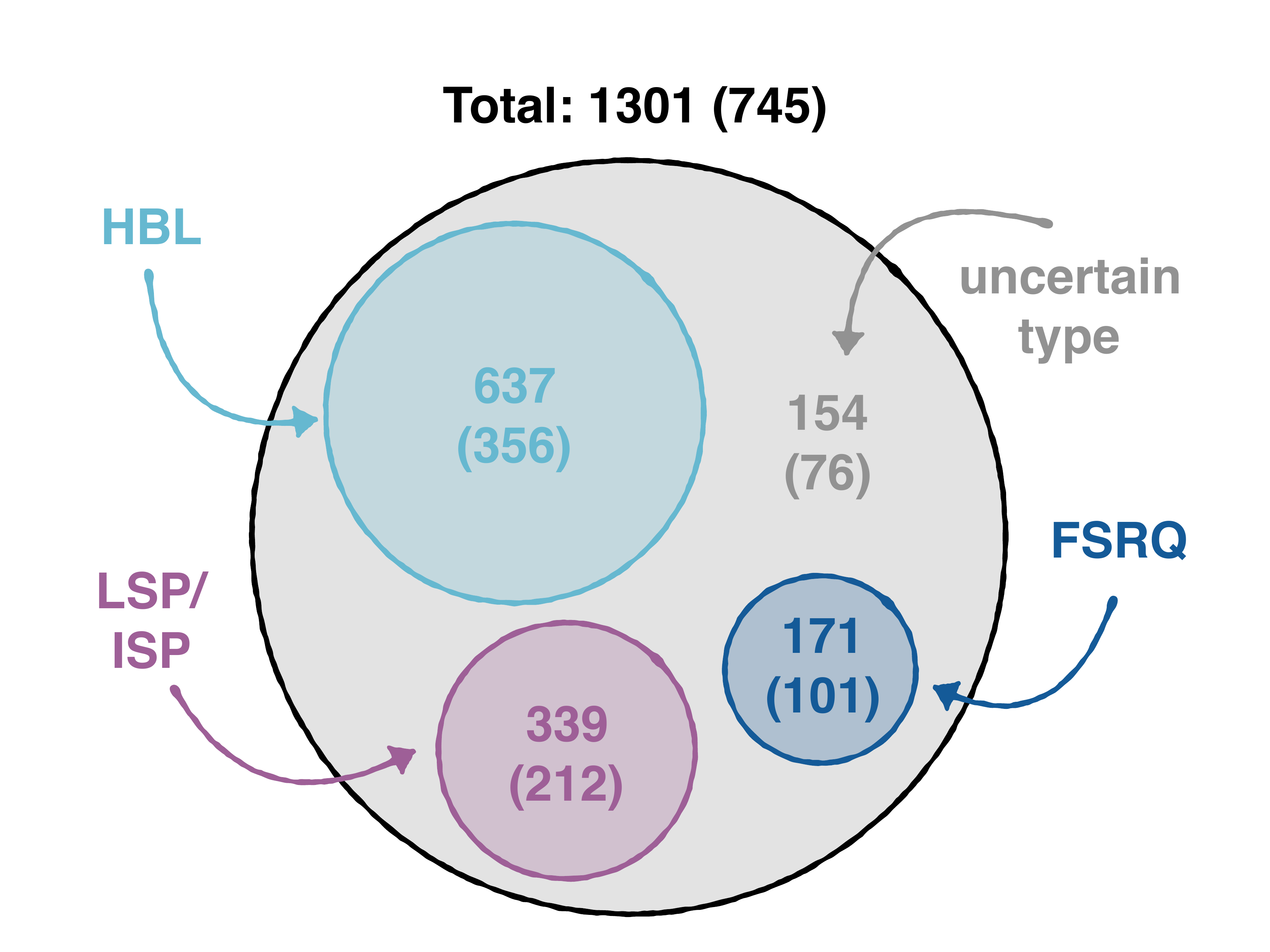}
\end{subfigure}%
\begin{subfigure}{.55\textwidth}
  \centering
  \includegraphics[width=1.\linewidth]{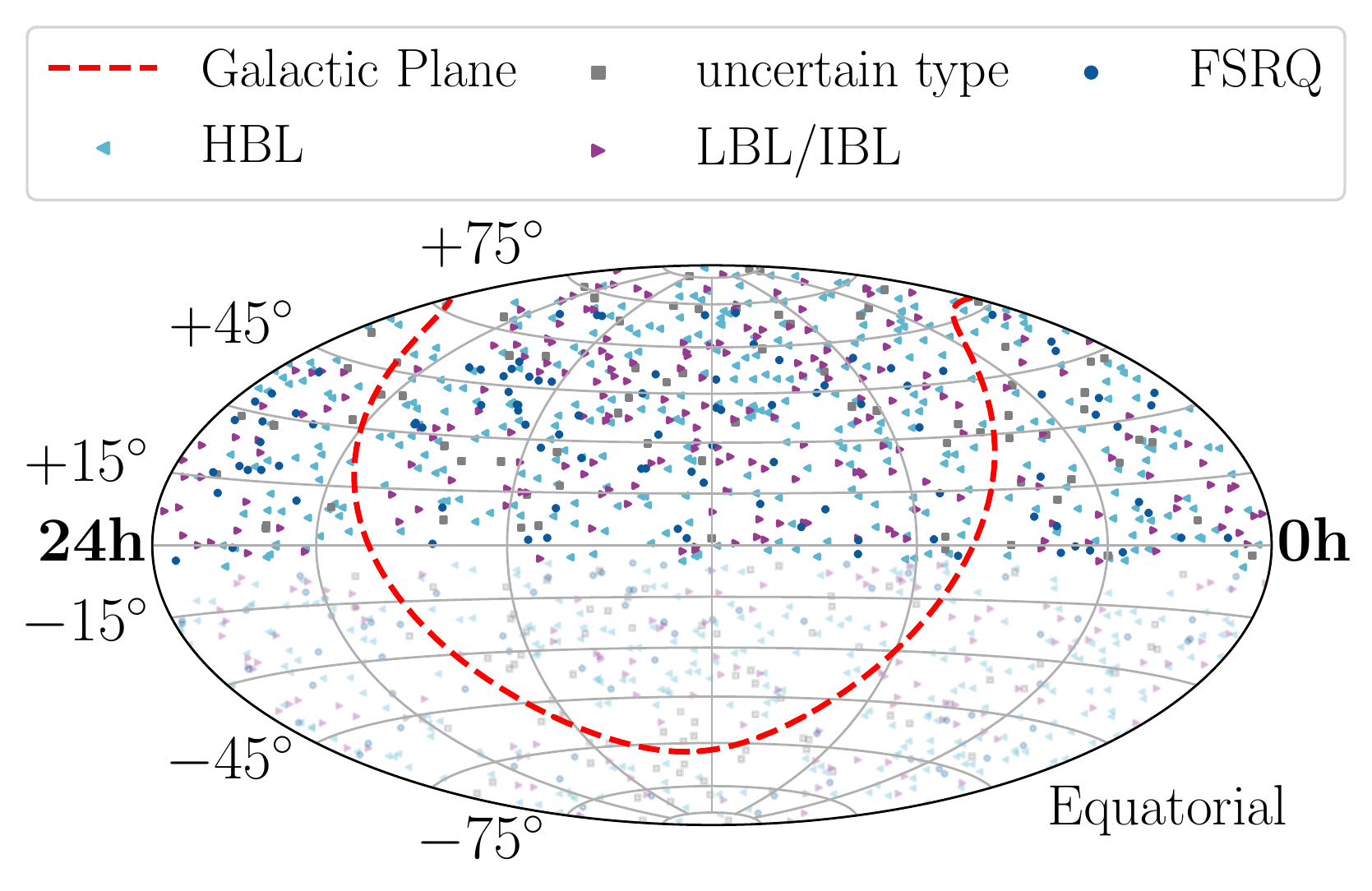}
\end{subfigure}
\caption{\textbf{Left:} Partition of the blazars from the 3FHL catalog into different categories. The     numbers show the frequency of the respective type (numbers in brackets only take Northern heimisphere sources, $\delta>-\SI{5}{\degree}$ into account). \textbf{Right:} Location of the 3FHL blazars. Southern hemisphere sources are shown with higher transparency.}
\label{fig:catalog}
\end{figure}

Motivated by previous studies that suggest a possible association between HSP BL Lacs (HBLs) and high-energy neutrinos \cite{Padovani:2014bha, Padovani:2016wwn} in addition to the detection of neutrinos from the direction of the TXS 0506+056 \cite{IceCube:2018dnn,IceCube:2018cha}, blazars from the 3FHL catalog are analyzed in this work. Since the different types of blazars show distinctly different properties at different wavelengths, this catalog is further split into FSRQs, HBLs and non-HBL BL Lacs. The partition of the catalog as well as the positions of all sources are illustrated in Figure \ref{fig:catalog}. In the following the outcome of all three blazar subsamples as well as the whole sample will be discussed. 

\section{Analysis method}
\label{sec:method}

\subsection{Unbinned likelihood ratio stacking approach}
\label{subsec:llh}

The through-going muon data set used for this analysis is vastly dominated by track-like events
arising from atmospheric neutrinos. 
Hence in order to find a significant indication of neutrino point sources it is essential to separate any
astrophysical neutrino signal from this much larger atmospheric background.
The statistical method that is used here to test the connection between the populations of blazars and the IceCube data is based on an unbinned likelihood ratio maximization using
the likelihood function

\begin{align}
  \mathcal{L}(\mu_s,\gamma|\{x\})  &=\prod_{j}^{N_{sample}}\prod_{i\in
  j}\left[\frac{\mu_s^{j}}{N_{tot}^{j}}
  S^{j}(x_i|\theta_s)
        +\left(1-\frac{\mu_s^{j}}{N_{tot}^{j}}\right) B^{j}(x_i|\theta_b)\right],
      \label{eq:llh}
\end{align}
with $j$ indexing the samples $j\in\{\textrm{IC59, IC79, IC86 I-VI}\}$ and
$i$ the respective event index. This likelihood is defined as a function of the spectral index
$\gamma$ and the mean number of expected signal events $\mu_s$, where $\mu_s^j$ describes the portion of expected signal events coming from sample $j$.
The separation power between background and signal likeliness of a specific scenario in Equation
\eqref{eq:llh} is mainly based on two event criteria. 
These criteria are the direction $\vec{n}_i$, including its uncertainty estimate $\sigma_i$ and the energy $E_i$ of the events. These individual event properties can be summarized as $x_i =\{\vec{n_i}, \sigma_i, E_i\}$. The physics parameters of a possible source are summarized in $\theta_s =\{\vec{n_s},\gamma\}$ while the background is characterized by $\theta_b$.

In general, neutrino signal events from point-like sources are expected to be distributed close
to the direction of their origin assuming a Gaussian-distributed spatial signal pdf 
for each position.
On the other hand events resulting from atmospheric background are spread locally uniform in
space, in the sense that the frequency of their appearance only depends on the detection efficiency of IceCube. 
In addition to the event position relative to the source the respective energy information $E_i$ can
be used to discriminate sources with a $E^{-\gamma}$ spectrum from the atmospheric background, which
approximately follows a very soft $E^{-3.7}$ spectrum.

Using the likelihood expression from Equation \eqref{eq:llh} the test statistic $\Lambda$ can be defined as
\begin{align}
          \Lambda &= -2\log \left[\frac{\mathcal{L}(\mu_s=0)}{\textrm{max}_{\mu_s,\gamma}\mathcal{L}(\mu_s,
          \gamma)} \right]. 
       \label{eq:TS}
\end{align}
By maximizing this likelihood ratio with respect to the mean number of expected signal events $\hat{\mu}_s$ and a globally fixed spectral index $\hat{\gamma}$ the validity of the point source hypothesis can be examined.

In case of a stacking analysis not only one point source but multiple point sources are assumed to
generate a cumulative astrophysical neutrino signal inside the detector. In order to test such a
scenario the signal pdf in Equation \eqref{eq:llh} has to be extended to the more general case of $M$ source candidates
\begin{align}
      S(x_i|\theta_s) &\rightarrow S_i^{stack}(x_i|\theta_s^{stack}) = \sum_{k=1}^M S(x_i|\theta_{s,k})  \cdot P(\theta_{s,k}|\theta_s^{stack}). 
       \label{eq:signal_stacking}
\end{align}
For simplification the sample index $j$ is omitted here.
The physics parameter of the ensemble of possible sources are characterized now by $\theta_s^{stack} = \{\{\vec{n}_{s,k}\}_k,\gamma\}$ with $\vec{n}_{s,k}$ indicating the position of source $k$.
This extended signal pdf can be regarded as superposition of the signal pdfs from the individual
sources. The relative source weights $P(\theta_{s,k}|\theta_s^{stack}) = W^k R^k(\theta_{s,k},\gamma) / \left(\sum_{l=1}^M W^l R^l(\theta_{s,l},\gamma)\right)$ consist of a detector specific weight $R^k(\theta_{s,k})$ accounting for the declination and energy dependent detection efficiency of IceCube and a theoretical weight $W^k$ accounting for the relative source strengths at the surface of the Earth.

\subsection{Catalog scans}
\label{subsec:scans}

Within the likelihood ratio formalism described above and used in the following all blazars are assumed to produce a neutrino signal at Earth with the same strength, 
yielding $W^k=\frac{1}{M}$. 
Nevertheless we can still examine a possible correlation between the VHE gamma-ray flux and
a neutrino flux in this analysis without making strong assumptions on the individual weights $W_k$.
For this purpose all blazar samples are partitioned into cumulative subsets according to the respective integrated gamma-ray fluxes.
The exact partitioning is chosen such that the number of new sources is approximately constant, yielding 10 flux subsamples. For each of the subsamples a pre-trial p-value is evaluated. The final post-trial p-value is defined as the best pre-trial p-value after trial correction. 
This procedure is repeated for all four blazar categories mentioned above, yielding in
total four final p-values.

\section{Results}
\label{sec:results}

\begin{figure}[t]
\centering
\begin{subfigure}{.5\textwidth}
  \centering
  \includegraphics[width=1.\linewidth]{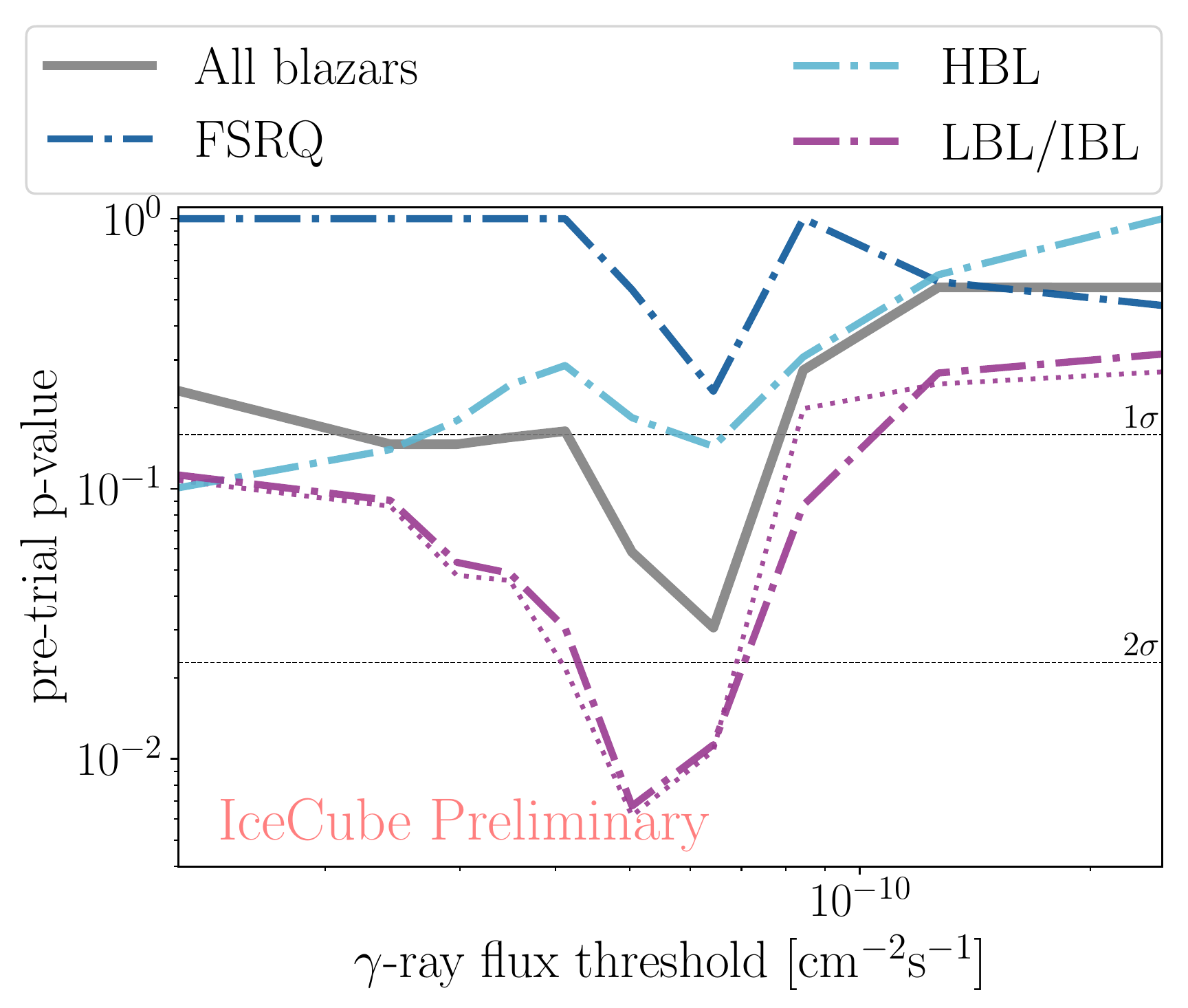}
\end{subfigure}%
\begin{subfigure}{.5\textwidth}
  \centering
  \includegraphics[width=1.\linewidth]{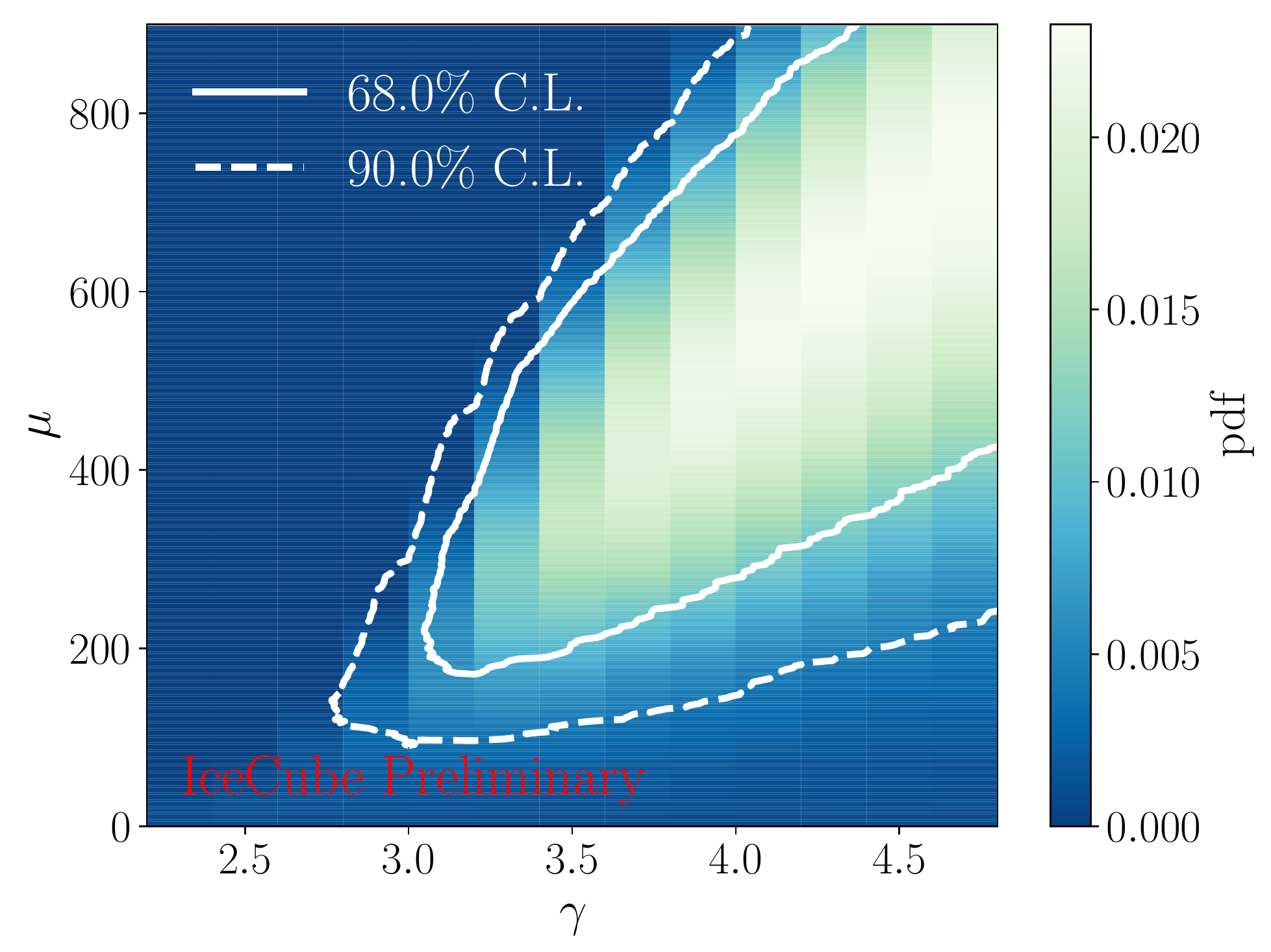}
\end{subfigure}
\caption{\textbf{Left:} Pre-trial p-values for blazars from the 3FHL catalog for different gamma-ray flux thresholds. \textbf{Right:} Confidence level contours on the mean number of signal events $\mu$ and the spectral index $\gamma$ for the subsample of LBL and IBLs which show the highest significance in this analysis.}
\label{fig:results}
\end{figure}

None of the blazar categories tested showed any significant evidence for a neutrino signal above background expectations. The pre-trial corrected p-value distributions are shown in Figure \ref{fig:results}. 
From none of these distributions a clear correlation between the integrated gamma-ray flux and a potential neutrino signal is visible. Having a direct correlation between gamma-rays and high-energy neutrinos would imply that sources with the highest gamma-ray flux should on average dominate the p-value. Adding more and more sources with lower neutrino power should further improve the significance. For all blazars except from FSRQs we can see an increase in significance up to a certain gamma-ray flux threshold, which is followed by a drop in significance afterwards. Yet looking at the results it seems like the largest contribution to the most significant p-values does not necessarily arise from the blazars with the highest gamma-ray flux. Hence for all these categories a direct correlation between the gamma-ray flux and a potential high energy neutrino flux can neither be confirmed nor ruled out completely. For FSRQs basically no excess is visible at all.  
The final post-trial p-values\footnote{Note that since none of the outcomes show a significant evidence we do not account for a trial factor for testing four different blazar categories.} for each of the blazar categories are shown in Table \ref{results}. 

\begin{table}[t]
	\centering
\begin{tabular}{lcccc}
\hline \hline
    Sample  &    p-value                    &  $\gamma$ & $\cdot\SI{e-19}{}\phi_{\SI{100}{\tera\electronvolt}}^{\SI{90}{\percent}}$    &  $r_{\textrm{max}} [\SI{}{\percent}]$ \\
\hline
    HBL              &   0.31 (+.5$\sigma$)    &  2.00 & \SI{6.5}{}  &   10.6 - 14.6   \\
    LBL/IBL          &   0.03 (+1.9$\sigma$)   &  2.00 & \SI{4.5}{}  &   8.2 - 11.9   \\
    FSRQ             &  0.56 (+0$\sigma$)      &   2.00 & \SI{1.6}{}  &   2.9 - 3.8   \\
    \multirow{2}{*}{All blazars} &   \multirow{2}{*}{0.12 (+1.2$\sigma$)}   &   2.00 & \SI{7.2}{}  &  13.0 - 16.7    \\
    & & 2.19 & \SI{8.4}{} &  9.7 - 13.9\\
\hline \hline
\end{tabular}
\caption{Post-trial p-values for all tested blazar categories. The median \SI{90}{\percent} C.L. upper limits $\phi_{\SI{100}{\tera\electronvolt}}^{\SI{90}{\percent}}$ for a single unbroken power-law $\frac{\partial \phi}{\partial E} = \phi_{\SI{100}{\tera\electronvolt}} \left(\frac{E}{\SI{100}{\tera\electronvolt}}\right)^{-\gamma}$ as well as the maximal possible contribution $r_{\textrm{max}}$ to the diffuse astrophysical muon neutrino flux \cite{Haack:2017dxi} of the respective 3FHL sources are shown as well. The flux normalizations $\phi_{\SI{100}{\tera\electronvolt}}^{\SI{90}{\percent}}$ are given in units of $\left[\SI{}{\per\giga\electronvolt\per\square\centi\meter\per\second} \right]$. The assumed global spectral index of the neutrino signal is represented by $\gamma$.} 
\label{results}
\end{table}

The most significant outcome is found for all LBL and IBL objects having a gamma-ray flux $\gtrsim \SI{5e-11}{\per\square\centi\meter\per\second}$. Since TXS 0506+056
is part of the LBL and IBL subsample of the 3FHL catalog it seems obvious to check if the excess in this category is connected to TXS 0506+056. TXS 0506+056 is an extraordinary source with an integrated gamma-ray flux above \SI{10}{\giga\electronvolt} of $\sim \SI{4e-10}{\per\square\centi\meter\per\second}$ \cite{TheFermi-LAT:2017pvy}.
Hence, in this analysis this source appears already in the highest gamma-ray flux threshold bin. Since by far the largest excess appears only once many sources with lower gamma-ray flux are added to the tested population it does not look like TXS 0506+056 dominates the excess at first glance. To get a more reliable statement about TXS 0506+056, it is removed from the tested sample. The corresponding pre-trial p-value distribution for the LBL and IBL blazars without TXS 0506+056 is illustrated by the dotted line in the left panel of Figure \ref{fig:results}. Figure \ref{fig:results} illustrates that TXS 0506+056 hardly affects this distribution and for the most significant gamma-ray flux threshold even slightly worsens the p-value. 

In order to understand this behavior one can have a look at the predictions of the physical source population parameters. 
Assuming the neutrino flux is similar for all sources , the confidence level contours on the mean number of signal events $\mu$ and the global spectral index $\gamma$ for these blazars can be estimated. They are shown in the right part of Figure \ref{fig:results}. It is notable that under these assumptions the IceCube data prefer a very soft global spectrum for these sources, which is not what we would expect from most theoretical blazar models. The TXS 0506+056 blazar on the other hand showed a harder spectral shape than this blazar population. Since we assume a global spectrum for all blazars in this analysis, 
TXS 0506+056 does not have a very strong impact.

In summary, these results do not show any significant evidence for a steady neutrino signal from blazars from the 3FHL catalog. Moreover, the contours for the most significant blazars indicate a very soft spectral index which is in good agreement with background expectations.

\section{Conclusions}
\label{sec:conclusion}

\begin{figure}[t]
\centering
\begin{subfigure}{.5\textwidth}
  \centering
  \includegraphics[width=1.\linewidth]{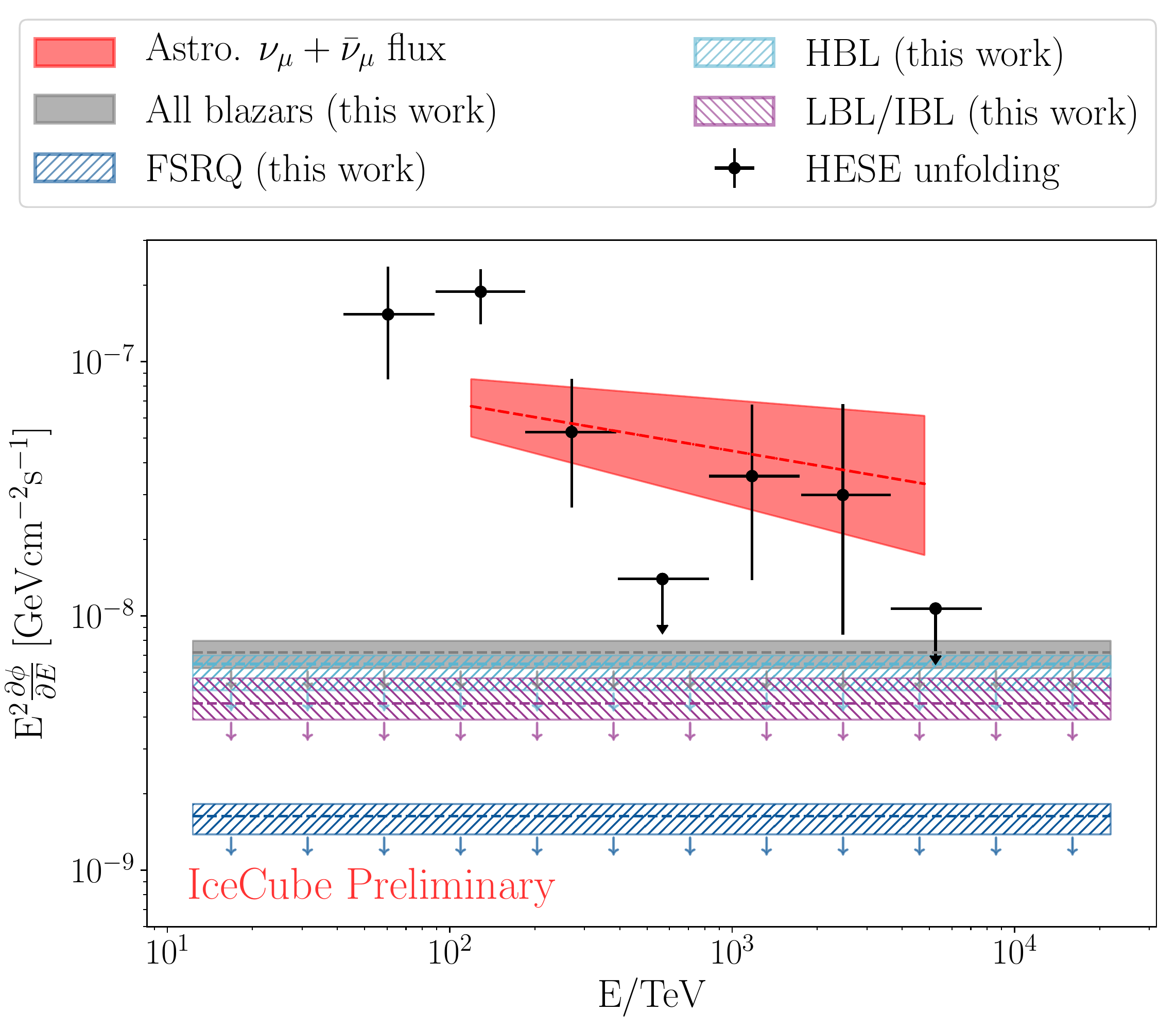}
\end{subfigure}%
\begin{subfigure}{.5\textwidth}
  \centering
  \includegraphics[width=1.\linewidth]{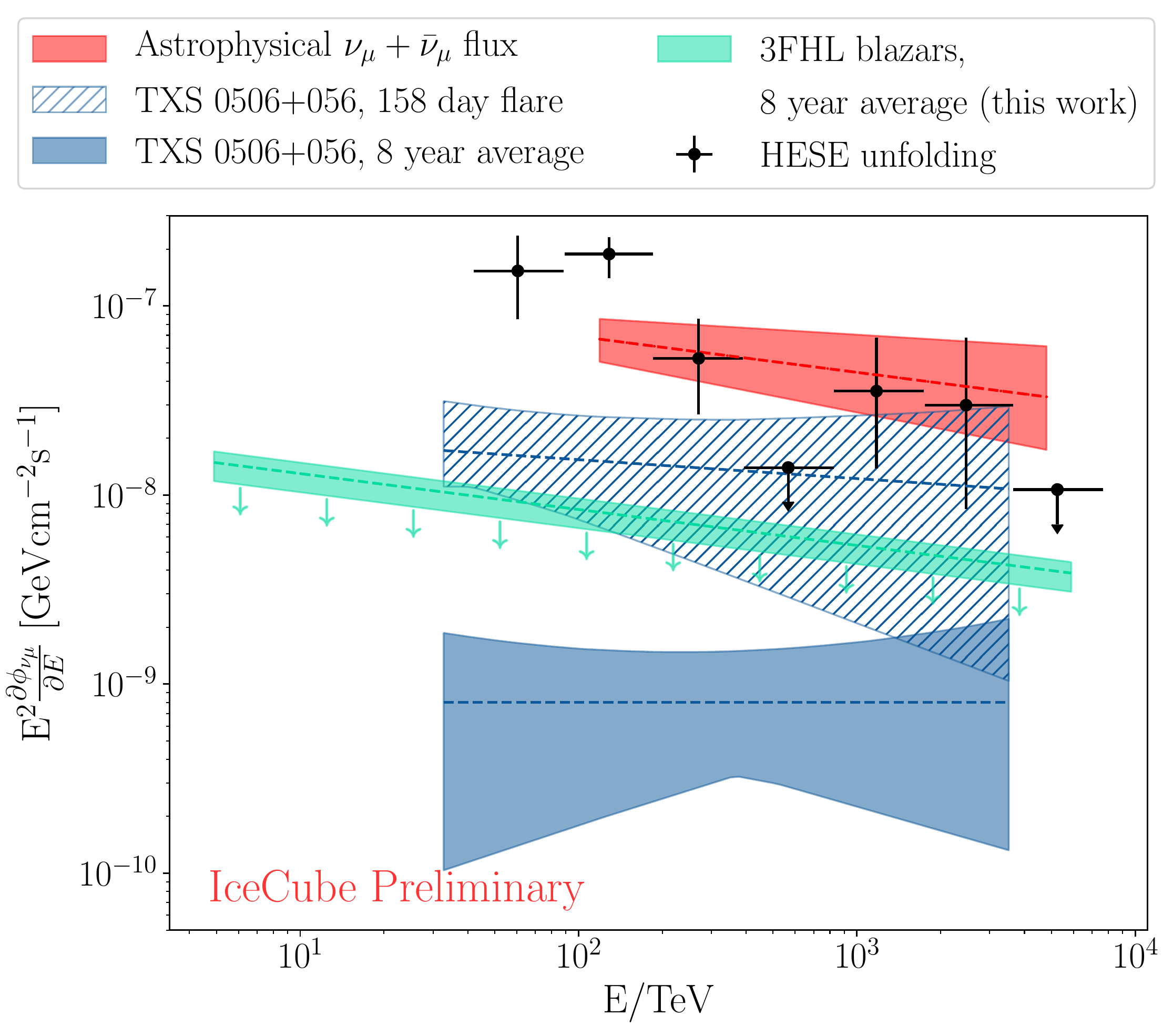}
\end{subfigure}
\caption{\textbf{Left:} \SI{90}{\percent} C.L. upper limits for blazar populations from the 3FHL catalog for $\gamma=2$. The dark-shaded bands illustrate the $1\sigma$ central band which arise from different weighting schemes for the individual sources. The energy ranges mark the \SI{90}{\percent} region where IceCube has the highest exclusion power for the
particular model. The best fit to the astrophysical
diffuse muon neutrino flux $\phi^{\textrm{astro}}_{\nu_{\mu} + \bar{\nu}_{\mu}}$ from
\cite{Haack:2017dxi, Claudio:2017dxi} including the \SI{68}{\percent} C.L. contours is displayed in red. \textbf{Right:} \SI{90}{\percent} C.L. limit on the average flux of all blazars from the 3FHL catalog for  $\gamma=2.19$. The blue shaded band shows the \SI{68}{\percent} C.L. band for the average muon neutrino flux of TXS, assuming that TXS 0506+056 is a steady neutrino source over $\sim 8$ years of measurement. The dashed band represents the average flux of this blazar during the \SI{158}{\day} flaring period in 2014/2015.
The data for both bands are taken from \cite{IceCube:2018cha}. 
Note that the HESE spectrum as well as the diffuse muon neutrino flux are integrated over the northern hemisphere
}
\label{fig:limits}
\end{figure}

Since no significant evidence for neutrino emission from the blazar samples analyzed was found, upper
limits on the $\nu_{\mu} + \bar{\nu}_{\mu}$-flux from these blazar samples were calculated. 
The \SI{90}{\percent} C.L. upper limit for the four tested blazar categories are shown in the left panel of Figure \ref{fig:limits} for a global unbroken power-law with a spectral index of $\gamma=2$.

The astrophysical muon neutrino flux observed by the IceCube collaboration yields a spectrum
compatible with a single unbroken power law and spectral index of $\gamma = 2.19 \pm 0.10$ between
\SI{119}{\tera\electronvolt} and \SI{4.8}{\peta\electronvolt} \cite{Haack:2017dxi}.
Under the assumption of a specific spectrum with spectral index $\gamma$, the maximum contribution of a blazar population to the diffuse flux can be evaluated. In this proceeding this contribution is evaluated as the ratio $r_{\textrm{max}} = \phi_E^{\textrm{cat}} / \phi_E^{\textrm{diff}}$ of the respective energy fluxes $\phi_E = \int E\cdot\frac{\partial \phi}{\partial E}dE$ between \SI{119}{\tera\electronvolt} and \SI{4.8}{\peta\electronvolt}. These fractions for all tested blazar categories are summarized in Table \ref{results}. 
Assuming a global spectral index of $\gamma=2$, blazars from the 3FHL catalog can not explain more than \SI{13.0}{\percent} to \SI{16.7}{\percent} of the average diffuse astrophysical energy flux. 
Assuming a similar spectrum as the one of the astrophysical neutrino flux, the maximal possible contribution from all 3FHL blazars to the diffuse flux can be determined to \SI{9.7}{\percent} to \SI{13.9}{\percent}
In summary, one can state that blazars from the 3FHL catalog can not explain the entire diffuse astrophysical muon neutrino flux, yet depending on the assumed spectral shape of the source population the maximal possible contribution can still vary a lot. 

Since TXS 0506+056 is part of the 3FHL blazar catalog it seems obvious to check if the outcomes from this analysis and the TXS 0506+056 studies are in agreement. The right panel of Figure \ref{fig:limits} shows average neutrino flux from the position of TXS 0506+056 for two different scenarios. The blue shaded band shows the \SI{68}{\percent} C.L. band for the average neutrino flux, assuming that TXS 0506+056 is a steady neutrino source over the lifetime of the IceCube detector measurements \cite{IceCube:2018cha}. On the other hand the dashed band represents the average flux scenario in which this blazar can be interpreted as a flaring source, producing neutrinos only within a \SI{158}{\day} flare (2014/2015) \cite{IceCube:2018cha}. In addition to these fluxes the \SI{90}{\percent} C.L. limits for all blazars from the 3FHL catalog are shown as well. These limits are again produced under the assumption that the whole blazar population consists of steady sources following a single power-law with a spectral index, $\gamma=2.19$. 
During the flaring period of TXS 0506+056 the average flux is on average higher than the \SI{90}{\percent} C.L. limit of the whole 3FHL blazar population. On the other hand, once the average neutrino activity of TXS 0506+056 over the whole 8 years of IceCube exposure, is compared to the limits from the 3FHL blazars one can see that on average TXS 0506+056 is much lower. Since these limits are only giving rejection power for the average neutrino emission of steady sources it is not absolutely surprising that individual sources can outshine these limits over shorter time periods. The neutrino observations of TXS 0506+056 and the limits presented in this work show very good agreement but also emphasize how exceptional the neutrino flare of TXS 0506+056 in 2014/2015 was.

\begin{spacing}{0.9}

\bibliographystyle{ICRC}
\bibliography{references}

\providecommand{\href}[2]{#2}\begingroup\raggedright\begin{thebibliography}{10}

\bibitem{IceCube:2018dnn}
{\bf IceCube, Fermi-LAT, MAGIC, AGILE, ASAS-SN, HAWC, H.E.S.S., INTEGRAL,
  Kanata, Kiso, Kapteyn, Liverpool Telescope, Subaru, Swift NuSTAR, VERITAS,
  VLA/17B-403} Collaboration, M.~G. Aartsen et~al., {\em Science} {\bf 361}
  (2018) eaat1378.

\bibitem{IceCube:2018cha}
{\bf IceCube} Collaboration, M.~G. Aartsen et~al., {\em Science} {\bf 361}
  (2018) 147--151.

\bibitem{Urry:1995mg}
C.~M. Urry and P.~Padovani, {\em Publ. Astron. Soc. Pac.} {\bf 107} (1995) 803.

\bibitem{Padovani:2014bha}
P.~Padovani and E.~Resconi, {\em Mon. Not. Roy. Astron. Soc.} {\bf 443} (2014)
  474--484.

\bibitem{TheFermi-LAT:2017pvy}
{\bf Fermi-LAT} Collaboration, M.~Ajello et~al., {\em Astrophys. J. Suppl.}
  {\bf 232} (2017) 18.

\bibitem{Aartsen:2016nxy}
{\bf IceCube} Collaboration, M.~G. Aartsen et~al., {\em JINST} {\bf 12} (2017)
  P03012.

\bibitem{Aartsen:2016oji}
{\bf IceCube} Collaboration, M.~G. Aartsen et~al., {\em Astrophys. J.} {\bf
  835} (2017) 151.

\bibitem{PhysRevD.78.034013}
S.~R. Kelner and F.~A. Aharonian, {\em Phys. Rev. D} {\bf 78} (Aug, 2008)
  034013.

\bibitem{Padovani:2016wwn}
P.~Padovani, E.~Resconi, P.~Giommi, B.~Arsioli, and Y.~L. Chang, {\em Mon. Not.
  Roy. Astron. Soc.} {\bf 457} (2016) 3582--3592.

\bibitem{Haack:2017dxi}
{\bf IceCube} Collaboration,  \pos{PoS(ICRC2017)1005}  (these proceedings).

\bibitem{Claudio:2017dxi}
{\bf IceCube} Collaboration,  \pos{PoS(ICRC2017)981}  (these proceedings).

\end{thebibliography}\endgroup
\end{spacing}
%

\end{document}